\documentclass[12pt]{iopart}
\usepackage{graphicx}

\begin{document}

\title{Quantum Dot coupled to a normal and a superconducting lead}
\author{MR Gr{\"a}ber, T Nussbaumer, W Belzig and C Sch{\"o}nenberger}
\address{Institut f{\"u}r Physik, Universit{\"a}t Basel,
Klingelbergstrasse 82, CH-4056 Basel, Switzerland}
\ead{christian.schoenenberger@unibas.ch}

\begin{abstract}
We report on electrical transport measurements in a carbon
nanotube quantum dot coupled to a normal and a superconducting
lead. Depending on the ratio of Kondo temperature $T_{K}$ and
superconducting gap $\Delta$ the zero bias conductance resonance
either is split into two side-peaks or persists. We also compare
our data with a simple model of a resonant level - superconductor
interface.
\end{abstract}

At low temperatures carbon nanotubes act as quantum dots.
Different transport regimes, depending on the transparency of the
contacts, such as Coulomb blockade and Kondo effect can be
realized~\cite{nygard,buit1}. Recently it has also been possible
to couple a carbon nanotube quantum dot in the Kondo regime to
superconducting leads, demonstrating a rich interplay of these two
many particle phenomena~\cite{buit2}. In this article we consider
a slightly different geometry, namely a quantum dot connected to
both a normal and a superconducting lead. These hybrid systems are
interesting for two reasons. First, the interplay of Kondo effect
and superconductivity can be examined on a different basis.
Various predictions have been made for this scenario, e.g.
suppression or enhancement of the conductance~\cite{cuevasSN},
side-peaks at the position of the superconducting
gap~\cite{clerkSN} and excess Kondo resonances~\cite{excess}.
Second, the structure mentioned above is the basic building block
of proposed Andreev entanglers making use of either the
0-dimensional quantum dot charging energy $U_{C}$
~\cite{RecherDot} or the 1-dimensional Luttinger repulsion energy
of a nanotube in order to spatially separate pairs of entangled
electrons~\cite{NTentangler}. In the following we will focus on
the interplay of Kondo effect and the superconducting lead.

\begin{figure}[h]
\begin{center}
\includegraphics [width=0.8\textwidth]{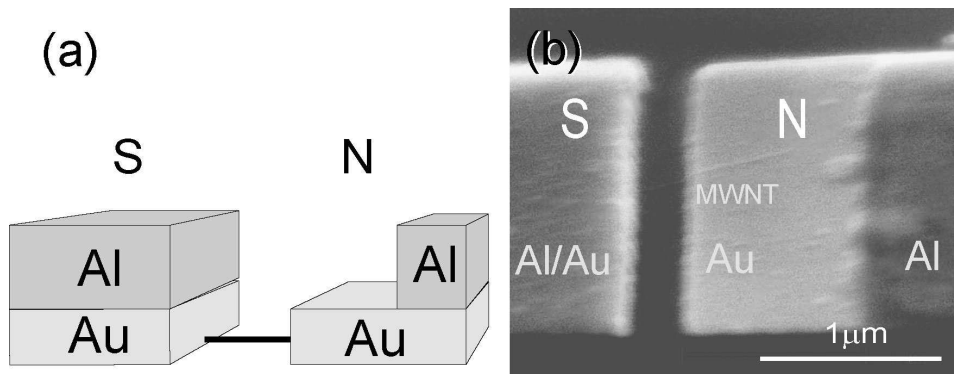}
\caption{(a) Schematics of the device. (b) SEM micrograph of the
sample.}
\label{Schematics}
\end{center}
\end{figure}

Here we report on electrical transport measurements of a Multi
Wall Carbon Nanotube (MWNT) quantum dot connected to a normal and
a superconducting lead. The sample is prepared as follows. First
MWNTs are spread on a degenerately doped silicon substrate, in the
experiment serving as a backgate, which is covered by a 400 nm
insulating layer of SiO$_{2}$. Then single nanotubes are contacted
by means of standard electron-beam-lithography and
e-gun-evaporation. Similar to reference~\cite{buit2} the
superconducting contact is a 45 nm Au/ 160 nm Al proximity
bilayer. However, by using tilt-angle-evaporation for the Al layer
one obtains a structure such as the one sketched in
figure~\ref{Schematics}(a). Whereas the left-hand side of the MWNT
is coupled to the superconducting Au/Al bilayer, the right-hand
electrode is formed simply by the 45 nm gold layer. There will
also be Al deposited on this side, but the spatial separation of
the nanotube-gold-contact and the Al film is fairly long
(approximately 1 $\mu$m). To check quantitatively whether also on
this side of the sample proximity effects have to be taken into
account one can estimate the Thoules energy \cite{Gueron}. The
Thoules energy represents an upper limit in energy for observing
superconducting correlations (assuming perfect barriers). One
obtains \mbox{$E_{T}=\hbar D/L^{2} \approx 3$\,$\mu$eV $\approx
10$\,mK} using a gold diffusion constant $D=5 \times 10^{-3}$
m$^{2}$/s (corresponding to an estimated Au mean free path of 10
nm) and a spatial separation L $\approx$ 1 $\mu$m. The experiment
is performed at \mbox{$90$\,mK}, hence, $kT$ is bigger than the
estimated $E_{T}$ and any proximity induced superconductivity on
the right sample contact can be safely neglected. Consequently the
sample geometry represents a S-QD-N structure. Figure \ref
{Schematics}(b) shows an SEM (Scanning Electron Microscope)
micrograph of the sample. Electrical transport measurements were
performed in a Kelvinox dilution refrigerator.

\begin{figure}[h]
\begin{center}
\includegraphics [width=0.8\textwidth]{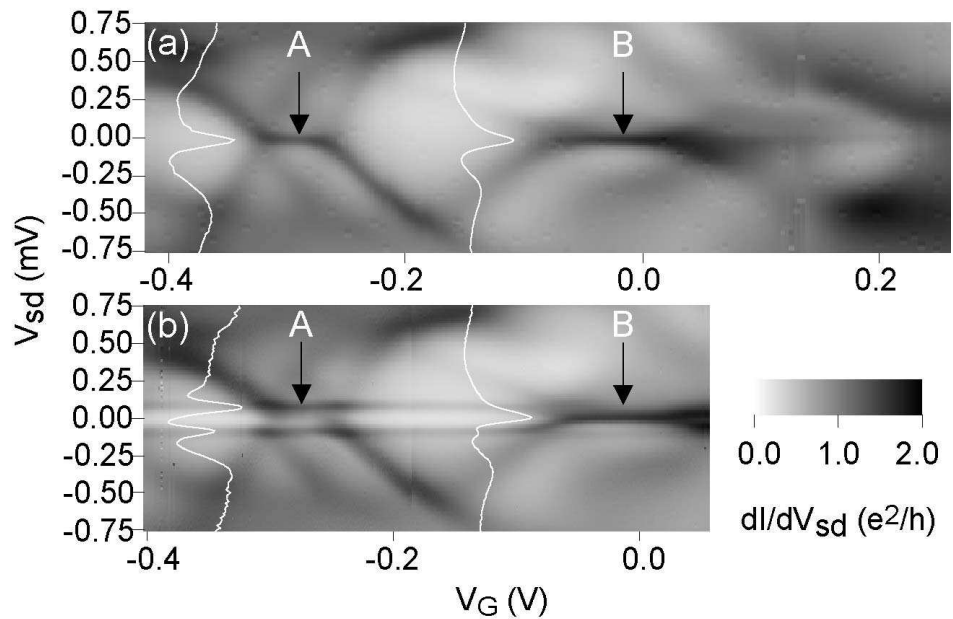}
\caption{(a) Grayscale representation of the normal state
conductance at 90 mK and B=25 mT (dark = more conductive). The
white curve on the left (right) shows the differential conductance
versus the applied source-drain voltage at the position of the
left (right) arrow. The two Kondo ridges are labelled ``A'' and
``B''. (b) Grayscale representation of the conductance in the
superconducting state at 90 mK and B=0 mT.} \label{Grayscales}
\end{center}
\end{figure}

By applying a small perpendicular magnetic field of $25$\,mT the
superconducting electrode is driven into the normal state and the
sample can be characterized in the N-QD-N configuration. This is
possible because the magnetic
 field is bigger than the Aluminum critical field but still small in
 terms of the Zeeman shift of the nanotube energy levels
 ($ E_{Zeeman}=g \mu_{B} B $ where g $\approx$ 2 is the
gyromagnetic ratio and $\mu_{B}$ the Bohr magneton). Figure
\ref{Grayscales}(a) shows a greyscale representation of the
differential conductance through the device at $T=90$ mK and
$B=25$ mT with varying backgate and source-drain voltage. Despite
some degree of disorder clear signs of Coulomb blockade diamonds
and the Kondo effect as manifest in the high conductance ridges
at zero bias voltage labelled ``A'' and ``B'' are visible. From the
size of the diamonds one can deduce a charging energy $U_{C}=
e^{2}/2C \approx$ 0.3 meV and a level spacing energy $\Delta
E\approx$ 0.3 meV. The coupling $C/C_{Gate}$ is of order 250. The
Kondo effect occurs when the number of electrons on the dot is odd
and it thus acts as a localized magnetic moment with spin 1/2.
Below the Kondo temperature $T_{K}$ the spins of the leads try to
screen the localized spin, i.e.  change its spin expectation
value to zero. In quantum dots this happens via fast spin-flip
processes allowed only on a short timescale within the Heisenberg
uncertainty principle. As a result of these processes between each
lead and the dot a resonance of the dot spectral density at the
chemical potential of the lead occurs which finally also causes a
resonance of conductance at zero bias. In the so-called unitary
limit for $T<<T_{K}$ a perfectly transmitting transport channel
opens up and the many-particle phenomenon Kondo effect reduces
effectively  to
a completely non-interacting problem~\cite{kondophysworld}.

\begin{figure}[h]
\begin{center}
\includegraphics [width=0.8\textwidth]{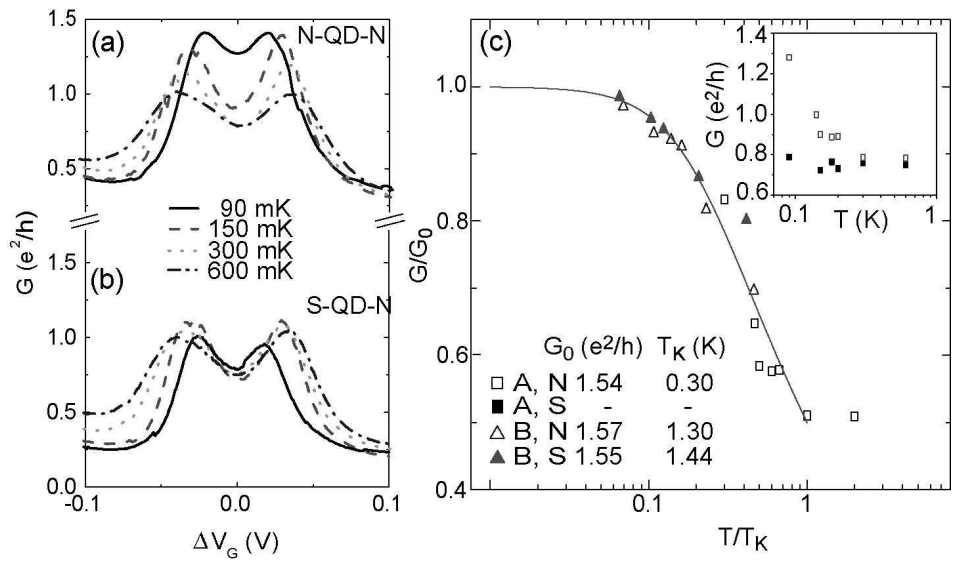}
\caption{(a) Linear response conduction of ridge ``A'' for
different temperatures in the normal state (B=25 mT). Labels
indicate the temperature in mK. (b) Like (a), but in the
superconducting state (B=0 mT). (c) Scaling plot of the maximum
Kondo conductance for ridge ``A'' and ``B'' in the normal and
ridge ``B'' in the superconducting state. The inset shows the
temperature dependence of the conductance at the center of ridge
``A'' in the normal (upper data) and superconducting (lower data)
state.} \label{Scaling}
\end{center}
\end{figure}

Figure \ref{Scaling}(a) shows the temperature dependence of the
linear conductance of ridge ``A'' for various temperatures in the normal state.
$T_{K}$ can be determined by examining the temperature dependence
of the linear conductance $G(T)$ on the Kondo ridge, i.e. exactly
in the middle of the two adjacent Coulomb peaks. As for the
classical Kondo effect one finds a logarithmic temperature
dependence. In order to determine $T_{K}$ we used the empirical
relation $G(T)=G_{0}/(1+(2^{1/s}-1)(T/T_{K})^2)^s$ where s=0.22
for a spin 1/2 system and the maximum conductance $G_{0} = 2 \:
e^{2}/h$ in the case of symmetric
coupling~\cite{Goldhaber-Gordon}. Best fits to our data yield
$T_{K,A}$ = 0.3 K, $T_{K,B}$= 1.3 K, $G_{0,A} = 1.54 \: e^{2}/h$
and $G_{0,B} = 1.57 \: e^{2}/h$. When plotting the normalized
conductance $G/G_{0}$ over the reduced temperature $T/T_{K}$ the
normal state data of ridges ``A'' and ``B'' collapse on a
universal locus, as seen in figure \ref{Scaling}(c). A further
rough estimate of the Kondo temperature is obtained by the width
of the resonant conductance peak yielding 0.6 meV ($\approx$ 0.72
K) and 0.1 meV ($\approx$ 1.2 K) for ridges ``A'' and ``B'',
respectively.

When one of the two electrodes enters the superconducting state
the Kondo effect is modified. As was shown in ~\cite{buit2} the
Kondo effect is suppressed by superconductivity only when the
superconducting gap $\Delta$ is bigger than $T_{K}$. A crossover
is expected for $\Delta\approx T_{K}$. However, in contrast to an
S-QD-S geometry here one faces an asymmetric situation and one has
to distinguish the nature of coupling between the dot and the
normal lead on one side and between the dot and the
superconducting lead on the other. Whereas the Kondo processes
between the normal lead and dot remain unaffected, two different
scenarios are possible for the superconducting lead-dot-coupling.
In a first case when $T_{K}$ is bigger than $\Delta$ one expects
the Kondo resonance to persist since quasiparticle states in the
superconducting electrode can participate in the Kondo spin-flip
processes. If, however, $T_{K}$ is smaller than $\Delta$ these
states will be missing and the Kondo coupling between the dot and
the superconducting lead will be strongly suppressed. Yet a
resonance of the dot spectral density with a renormalized Kondo
temperature (Kondo resonance width) $T_{K}^*<T_{K}$ remains, which is caused by the
Kondo processes between the normal lead and the dot. Whether one
actually sees an enhancement of zero-bias conductance at
temperatures below $T_{K}^*$ will now depend on the relevant dot
energy scales such as the charging energy $U_{C}$ (suppresses
Andreev reflections at the dot-superconductor interface) and the
coupling strength on both sides $\Gamma_{S}$ and $\Gamma_{N}$. In
certain parameter regimes it thus should also be possible to
enhance the conductance up to $4 \:e^{2}/h$, which is the maximum
value for a single perfectly-transmitting
channel~\cite{beenakker}.

Figure~\ref{Grayscales}(b) shows the conductance through our
device in a greyscale representation for the superconducting state
at T=90 mK and B=0 mT. The magnitude of the superconducting gap
can be deduced from the horizontal feature at
$V_{sd}=\Delta\approx$ 0.09 meV in good agreement with
\cite{buit2}, yielding a transition temperature $T_{C}\approx1$ K.
For the energy scales of our quantum dot we thus obtain $\Delta E
\approx U_{C} \approx 3\Delta$. We now focus on the two Kondo
regions in the superconducting case. In the case of ridge ``B''
with the width of the Kondo resonance being bigger than the
superconducting gap ($T_K/\Delta \approx 1.3$) both the greyscale
plot and the temperature dependence of the Kondo conductance
remain almost identical to the normal state, i.e. a strong
zero-bias conductance resonance and a logarithmic temperature
dependence at low temperatures. However, for temperatures
approaching the transition temperature $T_{C}$ the conductance in
the S-QD-N case is slightly higher than in the N-QD-N case,
similar to what one would expect for a channel with constant
transmission in the BTK model \cite{BTK}. When fitting the
temperature dependence with the same formula as above one obtains
a slightly enhanced Kondo temperature of $T_{K} = 1.44$ K and a
slightly reduced maximum conductance of $G_{0} = 1.55$ e$^{2}$/h.
For this fit we only considered temperatures sufficiently below
$T_{C}$ in order to exclude the BTK-like conductance enhancement
mentioned above. The data also collapse on the universal Kondo
locus, see figure \ref{Scaling}(c).

The resonance of ridge ``B'' remains in the superconducting state,
but its conductance is not increased. At first sight this behavior
seems surprising, since resonances indicate a high effective
transmission for which a doubling in conductance is expected in
the unitary limit. This, however, only holds for a symmetrically
coupled junction. Our observation is in quantitative agreement
with the theoretically expected conductance if we account for the
asymmetry. We consider the unitary limit for which the results for
non-interacting electrons should hold. The maximum conductance (at
resonance) of a transport channel between two normal electrodes is
given by
$G_{0}=(2\,e^{2}/h)\;4\Gamma_{L}\Gamma_{R}/(\Gamma_{L}+\Gamma_{R})^2$.
From our data we obtain $G_{0}=1.57\;e^{2}/h$ and thus a relative
asymmetry of the lead coupling of $\Gamma_{L}/\Gamma_{R}=0.37$ (or
the inverse). Between a normal and a superconducting lead the
maximum Andreev conductance has the form
$G_{0}=(4\,e^{2}/h)\;[2\Gamma_{L}\Gamma_{R}/(\Gamma_{L}^2+\Gamma_{R}^2)]^2$~\cite{beenakker}.
Using the $\Gamma$-ratio determined before one obtains for the
resonance conductance in the superconducting state
$G_{0}=1.69\;e^{2}/h$. This value is only slightly higher than the
one in the normal state and therefore explains our experimental
observation.

\begin{figure}[h]
\begin{center}
\includegraphics [width=0.8\textwidth]{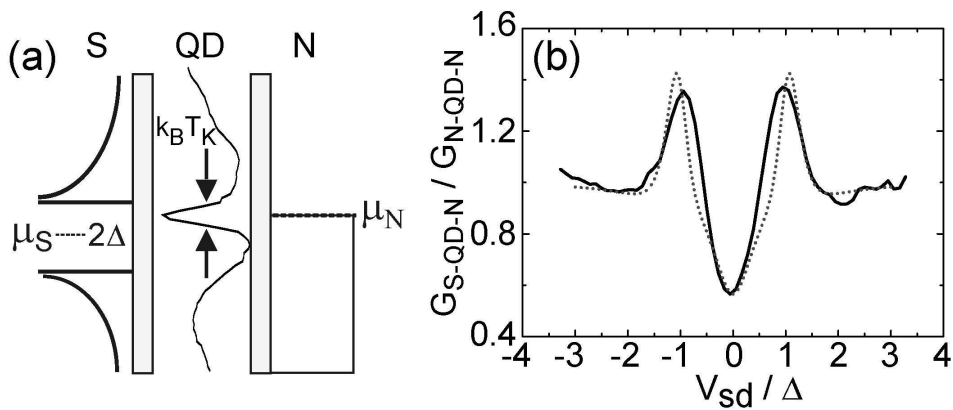}
\caption{(a) Simplified schematics of a quantum dot coupled to a
normal and a superconducting lead in the Kondo regime. (b) Solid
line: Superconducting conductance of ridge ``A'' versus
source-drain voltage at 90 mK normalized by the normal state
conductance. Dashed line: Simulation with the parameters given in
the text.} \label{theory}
\end{center}
\end{figure}

In case of ridge ``A'' the scenario is different ($T_K/\Delta
\approx 0.3$). The superconducting electrode results in a
suppression of the zero bias Kondo conductance enhancement but
high conducting side ridges at the position of the gap occur. This
can be understood when taking a look at figure \ref{theory}(a)
where the electronic spectrum of the quantum dot and of the leads
is depicted. The remaining Kondo coupling between the normal lead
and the dot results in a resonance of the dot spectral density
pinned to the normal lead chemical potential $\mu_{N}$. At a bias
of $V_{sd}=\Delta$ the superconductor quasiparticle spectrum and
the normal lead chemical potential (and thus the Kondo resonant
level) are lined up and resonant transport occurs. The linear
conduction of ridge ``A'' in the superconducting state versus gate
voltage is shown for various temperatures in figure
\ref{Scaling}(b). In the inset of figure \ref{Scaling}(c) the
conductance at the center of ridge ``A'' is plotted versus
temperature for both the normal and the superconducting state.
Whereas there is an increase of conductance below
$T_{K}\approx0.3$ K for the normal state data, the conductance
remains more or less constant in the superconducting case. Thus we
were not able to perform a fit in order to determine the
renormalized Kondo temperature $T_{K}^{*}$. However, we suspect
$T_{K}^{*}$ to be of the order of 100 mK since the 90 mK data show
an increase of conductance. The temperature dependence of the
conductance on the $\Delta$-side-peaks (data not shown) does not
show logarithmic behavior down to our lowest temperatures either,
but similar to the linear response conductance an enhancement for
the 90 mK data. This might indicate Kondo coupling at non-zero
bias between the quasiparticles in the superconducting and the
normal lead.

The behavior of ridge ``A'' can find a simple explanation by
assuming a strong suppression of the Kondo coupling between the
dot and the superconducting lead lowering the effective
transmission $T_{eff}$ of this interface to values comparable to
tunnel barriers. The current through our device in the
superconducting state is then given by \cite{Tinkham}:
\begin{equation}
I=2e/h T_{eff}
\int_{-\infty}^{\infty}N_{Dot}(E)N_{S}(E+eV)(f(E)-f(E+eV))dE
\end{equation}
with $N_{S}(E)$ being the BCS density of states in the
superconducting lead, $N_{Dot}(E)= Re(i w(E+ i w)^{-1})$ the QD
local density of states of the resonant level with width $w$,
$f(E)$ the Fermi function and $T_{S}<<1$ the transmission of the
barrier. Similar to \cite{buit3} we included a broadening $\gamma$
of the BCS density of states which we attribute to the additional
Au layer separating tube and Al layer. In figure \ref{theory}(b)
we plot the conductance of ridge ``A'' in the superconducting
state normalized by that in the normal state from both our
experimental data and simulations. Best agreements are obtained
with $w=0.3\: \Delta$, $T_{eff}=0.15$ and $\gamma=0.05\: \Delta$.
Finite temperature is taken into account by setting T=0.1$\Delta$.
In the normal state we approximated the Kondo conductance peak as
a Lorentzian. Comparison with our experimental data yields for the
width of the Lorentzian $0.6\: \Delta$, a maximum conductance of
1.35 $e^2/h$ and a background conductance of 0.75 $e^2/h$. The
proposed model clearly reproduces the main features of the
experimental data, however precise quantitative agreement remains
difficult. A possible explanation of e.g. the bigger width of the
$\Delta$-peaks is an energy-dependent transmission matrix element
(which we assumed to be constant) of increasing magnitude as the
applied bias approaches $\Delta$ due to Kondo coupling between the
normal lead, the dot and the superconducting quasiparticles.

In this paper we studied a carbon nanotube quantum dot in the
Kondo regime coupled to a normal and a superconductor. In the case
of $T_{K}<\Delta$ the Kondo ridge at zero bias disappears and
peaks at the position of the gap occur. For this scenario we
proposed a simple tunnelling model explaining all significant
features of the conductance curves. In the case $T_{K}>\Delta$ the
Kondo resonance persists but does not show an enhancement of the
conductance compared to the normal state at the lowest
temperatures accessible in our experiment. Future experiments will
have to (a) clarify whether the Kondo resonance can actually be
enhanced in presence of the superconducting electrode by tuning
the coupling asymmetry $\Gamma_{S}/\Gamma_{N}$ and (b) explore the
possibility of generating pairs of entangled electrons by making
use of nanotubes coupled to normal and superconducting
leads~\cite{NTentangler}.

We thank M. Buitelaar, B. Choi, L. Grueter, T. Kontos and S. Sahoo
for experimental help and A. Clerk, J. Cuevas, T. Kontos, A. Levy
Yeyati and P. Recher for discussions. We thank L. Forro for the
MWNT material and J. Gobrecht for the oxidized Si substrates. This
work has been supported by the Swiss NFS and the NCCR on
Nanoscience.

\end{document}